\documentclass[twocolumn,
amsmath,amssymb,
aps,
prl,
]{revtex4-1}

\usepackage{amsmath}
\usepackage{amssymb}
\usepackage{graphicx}
\usepackage{dcolumn}
\usepackage{bm}
\usepackage{color}

\DeclareMathOperator{\Imag}{Im}
\DeclareMathOperator{\Real}{Re}

\newcommand{\E}{\textbf{E}}

\newcommand{\J}{\textbf{J}}

\newcommand{\Hv}{\textbf{H}}
\newcommand{\B}{\textbf{B}}

\newcommand{\rv}{\textbf{r}}
\newcommand{\vv}{\textbf{v}}

\newcommand{\na}{\nabla}

\begin{document}
	\title{Absence of unidirectionally propagating surface plasmon-polaritons in nonreciprocal plasmonics}
	
	\author{Siddharth Buddhiraju}
	\author{Yu Shi}
		\author{Alex Song} 	
		\author{Casey Wojcik}
		\author{Momchil Minkov}
				\author{Ian A.D. Williamson}
								\author{Avik Dutt} 	
	\author{Shanhui Fan}%
	\email{shanhui@stanford.edu}
	\affiliation{%
		Ginzton Laboratory, Department of Electrical Engineering, Stanford University, Stanford, CA
	}%
	\date{\today}

\begin{abstract}
In the presence of an external magnetic field, the surface plasmon polariton that exists at the metal-dielectric interface is believed to support a unidirectional frequency range near the surface plasmon frequency, where the surface plasmon polariton propagates along one but not the opposite direction. Recent works have pointed to some of the paradoxical consequences of such a unidirectional range, including in particular the violation of the time-bandwidth product constraint that should otherwise apply in general in static systems. Here we show that such a unidirectional frequency range is nonphysical, using both a general thermodynamic argument, and a detailed calculation based on a nonlocal hydrodynamic Drude model for the metal permittivity. Our calculation reveals that the surface plasmon-polariton remains bidirectional for all frequencies. This work overturns a long-held belief in nonreciprocal photonics, and highlights the importance of quantum plasmonic concepts for the understanding of nonreciprocal plasmonic effects.
\end{abstract}

\maketitle

In the past two decades, there have been significant developments in the field of plasmonics, which explores surface plasmon-polaritons that exist at the metal-dielectric interface for nanoscale control of light \citep{maierPlasmonicReview, ozbayPlasmonicReview, schullerPlasmonicReview}. Most plasmonic structures satisfy the Lorentz reciprocity theorem \cite{jalasWhatIs}. On the other hand, in the presence of external magnetic field, the behavior of surface plasmon-polaritons becomes nonreciprocal. Such nonreciprocal surface plasmon-polaritons have generated substantial interest \cite{belotelovNonreciprocalPlasmonics, temnovNonreciprocalPlasmonics, chinNonreciprocalPlasmonics, enghetaNonreciprocalPlasmonics} since they represent a fundamentally different regime of light propagation, having potential importance for applications such as sensing and information processing. 

A particularly significant effect of nonreciprocal plasmons is the existence of a unidirectional frequency range \citep{wallisSMP1, hartsteinSMP, zongfuSMP,huSMP,shenSMP,tsakmakidisSMP}. With the metal described by the Drude model, in the presence of an external magnetic field, there exists a frequency range where the surface plasmon-polariton can only propagate along one direction. In a recent paper \citep{tsakmakidisSMP}, it was noted that the existence of such unidirectional surface plasmon can lead to resonator structures that violate the time-bandwidth product constraint. Subsequently, Refs. \citep{tsang, aluTBP} argued that breaking the time-bandwidth product should not be possible based on a coupled-mode theory analysis, and that doing so may violate the second law of thermodynamics. Motivated by these considerations, it becomes important to re-examine the fundamental physical assumptions that give rise to unidirectionally propagating surface plasmon-polaritons.

Within the local Drude model, the existence of the unidirectional frequency range depends on the behavior of the model in the limit of large wavevectors. On the other hand, it has long been known \citep{ginzburg} that nonlocal effects become important in this limit. In this Letter, we show that there should not be such a unidirectional frequency range in the spectrum of the surface plasmon polariton once nonlocal effects are taken into account. Instead, there will always be propagating modes in both directions. We illustrate this by detailed calculations based on the hydrodynamic model for the metal dielectric function. We also present a general thermodynamic argument to show that the main conclusion of the paper, i.e., the absence of unidirectionally propagating surface plasmons-polaritons, should hold for any physical nonlocal model of dielectric function. 

\textit{Drude model.---}We start with a brief review of the dispersion relation of the surface plasmon-polariton at the metal-dielectric interface. Throughout the paper for simplicity we refer to any system with a strong plasmonic response as a metal. In addition to the usual free-electron metal, such a `metal' can also include heavily doped semiconductors which exhibit plasmonic response at the infrared wavelengths. Within the local Drude model, in the presence of a static magnetic field $B_0$ along the $-\hat{y}$ direction, the frequency($\omega$)-dependent dielectric function of metal has the form:
	\begin{align}
	\frac{\epsilon(\omega)}{\epsilon_\infty } &= 1- \frac{\omega_p^2}{(\omega+i\gamma)^2-\omega_c^2} \nonumber \\
	&\times \begin{pmatrix}
	1 + i\frac{1}{\omega\gamma} & 0 & i\frac{\omega_c}{\omega} \\ 0 & \frac{(\omega+i\gamma)^2-\omega_c^2}{\omega(\omega+i\gamma)} & 0 \\ -i\frac{\omega_c}{\omega} & 0 & 1+i\frac{1}{\omega\gamma}
	\end{pmatrix}, \label{eq:tensor}
	\end{align}
	where $\epsilon_\infty$ is the dielectric response from the bound electrons and ions, $\omega_p$ the plasma frequency, $\gamma$ the phenomenological loss rate and $\omega_c = eB_0/m$ the cyclotron frequency, with fundamental charge $e$ and effective mass of the free carriers $m$.
	 
\begin{figure}
		\centering
		\includegraphics[scale=0.5]{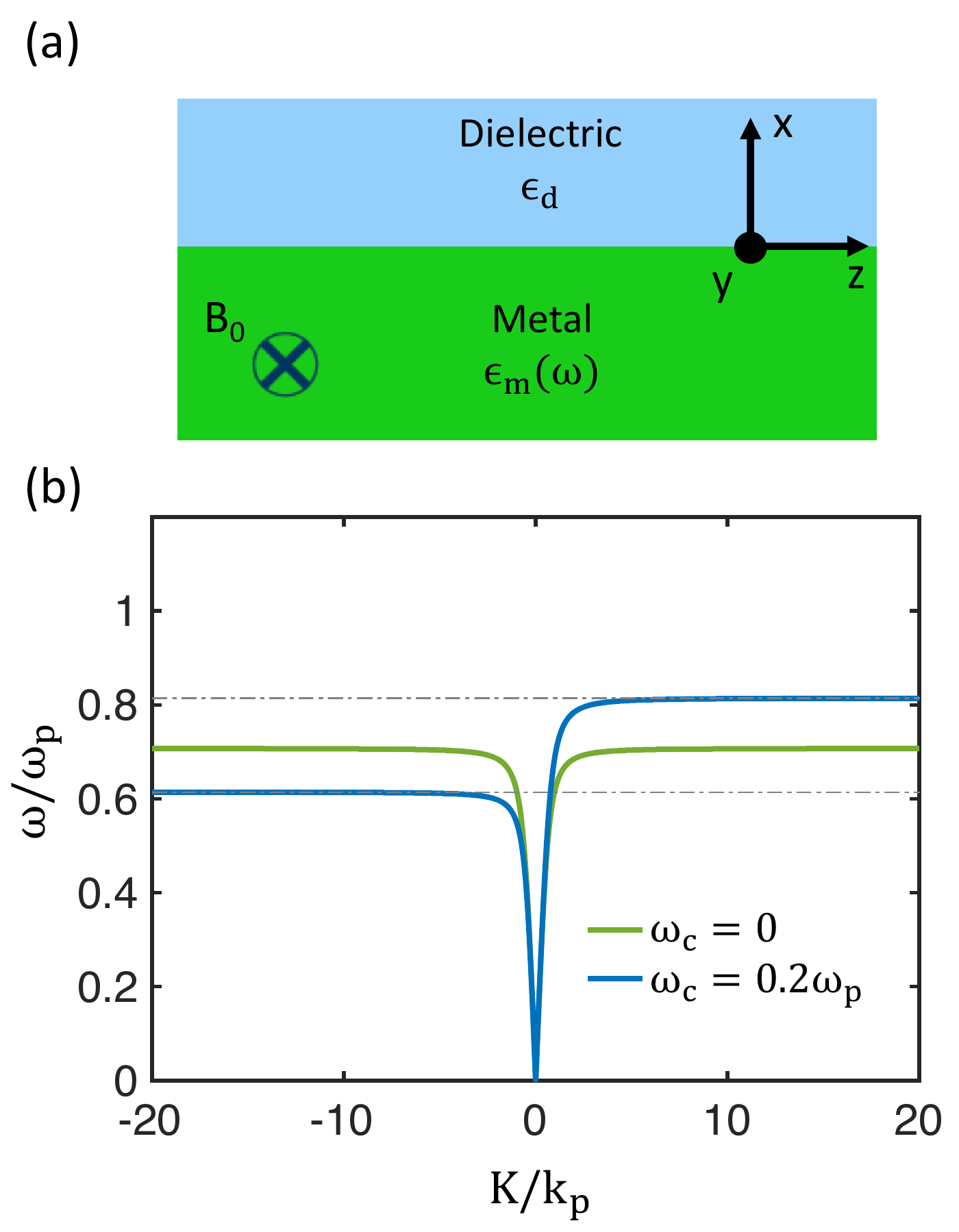}
		\caption{(a) An interface between a dielectric and a metal described by the Drude model. (b) Dispersion relation of surface plasmon-polariton propagation at the interface in the absence (green curve) and in the presence (blue curve) of an external magnetic field $B_0$ indicated in (a). $K$ is the wavevector component parallel to the interface, and $\omega_c=eB_0/m$ is the cyclotron frequency.}
		\label{fig:localmodel}
	\end{figure}

	Consider the metal-dielectric interface shown in Fig. \ref{fig:localmodel}(a). When $B_0 = 0$, in the near-lossless limit of $\gamma\rightarrow 0$, the dispersion relation $\omega(K)$ of the surface plasmon-polariton is shown by the green curve in Fig. \ref{fig:localmodel}(b), where $K$ is the wavevector parallel to the interface. Here we assume that $\epsilon_\infty=1$, and that the dielectric is air with $\epsilon_d = 1$. Since the system is reciprocal, we have $\omega(K) = \omega(-K)$. In the $K\rightarrow \pm\infty$ limit, the frequency of the surface plasmon-polariton approaches the surface plasmon frequency $\omega_{sp} = \omega_p/\sqrt{2}$. 
	
	When $B_0 \neq 0$, again assuming the near-lossless limit, the dispersion relation for the same interface system is shown by the blue curve in Fig. \ref{fig:localmodel}(b). Since the system is no longer reciprocal, we have $\omega(K) \neq \omega(-K)$. Moreover, the surface plasmon frequencies for the forward and backward directions are unequal at the $|K| \rightarrow\infty$ limit, opening a unidirectional frequency range around $\omega_{sp}$ where the surface plasmon-polariton propagates only along the positive-$K$ direction. The existence of such a unidirectional frequency range is the key to the unusual time-bandwidth product behavior reported in Ref. \citep{tsakmakidisSMP}. On the other hand, as observed from Fig. \ref{fig:localmodel}(b), the existence of such a unidirectional range is strongly dependent on the behavior of the Drude model in the $|K| \rightarrow\infty$ limit. And yet, it is well known that the local Drude model of Eq. \eqref{eq:tensor} is no longer adequate in this limit \citep{ginzburg}, and instead the spatially dispersive or `nonlocal' behavior of the electromagnetic response of the metal must be taken into account. Therefore, to understand the potential physics of such undirectional propagation, it is essential to take into account the effect of the nonlocal dielectric function.

\textit{Hydrodynamic Drude model.---}There exist many treatments of nonlocality, such as those based on the hydrodynamic model \cite{bennettHDM, razaHDM, ciraciHDM}, the random phase approximation \cite{feibelmanRPA, abajoNonlocal}, and a quantum corrected model \citep{estebanQuantumPlasmonics}. The description of plasmonic properties using these models is closely related to the emerging area of quantum plasmonics, where the quantum nature of the electron gas plays a significant role \citep{mortensenGNOR, MaierReview, jacobPlasmonicReview, estebanQuantumPlasmonics}. Here, we briefly discuss the hydrodynamic model, a simple analytical model that has often been used to describe nonlocal response in deep subwavelength metallic structures \cite{razaHDM, razaReview, toscanoHDM}, and recently in nanoparticles made of doped semiconductors such as InSb \cite{maack}. We refer readers to Refs. \citep{pitarkeReview, razaReview} and references therein for a detailed overview of nonlocality in surface plasmon-polaritons as well as a derivation of the hydrodynamic model. 

In this model, the collective motion of electrons is described using a density $n(\rv, t)$, a velocity $\vv(\rv,t)$, and an energy functional that can be appropriately chosen to describe the internal kinetic energy as well as interactions. We follow Ref. \citep{razaReview} to employ the Thomas-Fermi approximation for the energy functional. The equations of motion of the free carriers in this approximation are given by \citep{boardmanNonlocal, sarmaNonlocal}
	\begin{eqnarray}
	\frac{\partial \vv}{\partial t} + \gamma\vv +  (\vv\cdot\na)\vv &=& -\frac{e}{m}(\E + \vv\times\B) - \beta^2\frac{\na n}{n} \label{eq:v}\\
	\frac{\partial n}{\partial t} &=& -\na\cdot(n\vv) \label{eq:n}
	\end{eqnarray}
	where $\beta$ is the nonlocal parameter proportional to the Fermi velocity $v_F$ \cite{halevi},
	\begin{equation}
	\beta^2 = \frac{\frac{3}{5}\omega + \frac{1}{3}i\gamma}{\omega+i\gamma}v_F^2.
\end{equation}		
Linearizing Eqs. \eqref{eq:v} and \eqref{eq:n}, and using $\J = -en_0\vv$, where $n_0$ is the equilibrium electron density, a single equation in the frequency domain can be obtained for $\J$:
	\begin{equation}
	\beta^2\na(\na\cdot\J) + \omega(\omega+i\gamma)\J = -i\omega(\omega_p^2\epsilon_0\E-\frac{e}{m}\J\times\B_0) 	\label{eq:j}
	\end{equation}
	where $\B_0$ is an externally applied dc magnetic field. This equation is coupled with Maxwell's equations, written using the $\E$ field,
	\begin{equation}
	\na\times\na\times\E = \frac{\omega^2}{c^2}\epsilon_{\infty}\E - i\omega\mu_0\J.	\label{eq:e}
	\end{equation}
	Unlike in the local model, the presence of the nonlocal term $\beta^2\na(\na\cdot)$ in this model requires an additional boundary condition \citep{razaReview} to determine the dispersion relation. Neglecting the quantum mechanical effect of electron spill-out at the boundary of the free-electron gas with the dielectric, the additional boundary condition required to solve Eqs. \eqref{eq:j} and \eqref{eq:e} is
	\begin{equation}
	\J\cdot\hat{\textbf{n}}= 0,
	\end{equation}
	which has the effect of imposing an infinite potential well for the electron gas at the metal-dielectric boundary. 
	
\begin{figure}
	\centering
	\includegraphics[scale=0.4]{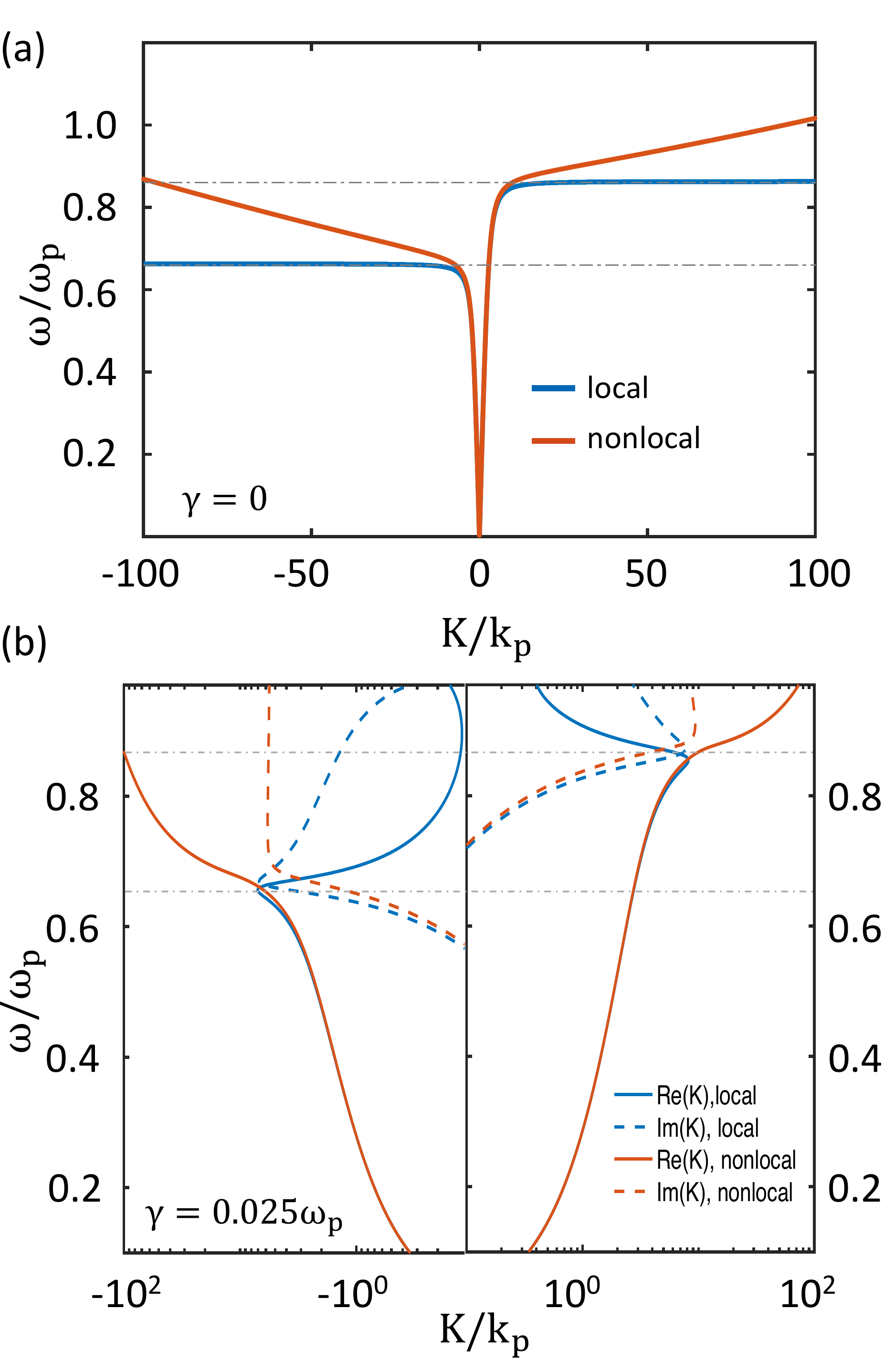}
	\caption{Dispersion relation for the interface considered in Ref. \citep{tsakmakidisSMP}, within the local (blue) and nonlocal (red) models. (a) A flat dispersion relation is obtained in the lossless ($\gamma=0$) local Drude model, resulting in a unidirectional gap indicated by the dotted grey lines. This gap is removed in the hydrodynamic Drude model with a high-$K$ counter-propagating wave. (b) Real (solid line) and imaginary (dotted line) parts of $K$ when $\gamma=0.025\omega_p$ in the local and the nonlocal models. $\Real(K)<\Imag(K)$ in the unidirectional gap in the local model, while $\Real(K)>\Imag(K)$ in the nonlocal model. Note that the $x-$axis is in log scale. $k_p = \omega_p/c$.}
\label{fig:nonlocalmodel}
\end{figure}	

To illustrate the effect of nonlocality on the nonreciprocal behavior of the surface plasmon-polaritons, we consider an interface where the dielectric layer is silicon ($\epsilon_d = 11.68$) and the metallic layer is n-doped indium antimonide (InSb), a material commonly used in demonstrating magneto-optical plasmonic effects\citep{hartsteinSMP,wallisSMP1,tsakmakidisSMP}. This interface was previously used in Ref. \citep{tsakmakidisSMP}, with the InSb layer treated using the local Drude model. The InSb layer has $\epsilon_\infty = 15.6$ and plasma frequency $\omega_p = 2\pi\cdot(2\times 10^{12}$ Hz). A constant dc magnetic field of $B_0 = 0.2$ T is applied in the $-\hat{y}$ direction to break reciprocity. Owing to a rather small conductivity effective mass for electrons, a large value\cite{maack} of $\beta = 1.07\times 10^6$ m/s is obtained at 300 K. Thus, the effect of nonlocality, which was not considered in Ref. \citep{tsakmakidisSMP}, is in fact prominent in the dielectric response of InSb.  In order to highlight the difference between the hydrodynamic model and the local Drude model, we first set $\gamma=0$. Using these parameters, we solve Eqs. \eqref{eq:j} and \eqref{eq:e} for surface-plasmon polariton dispersion relation at the Si-InSb interface. The red curve in Fig. \ref{fig:nonlocalmodel}(a) depicts the dispersion relation in the hydrodynamic model, and the blue curve in the local Drude model. The dispersion relations from the two models are almost the same for small $K$, but deviate as $K$ becomes larger. In particular, within the hydrodynamic model, there is no longer a unidirectional frequency range. At every frequency, there are both a propagating and a counter-propagating mode. We also note that the predictions between the local and nonlocal models start to deviate for K = 0.4 $\mu$m$^{-1}$. Thus, in this system, nonlocal effects become important even for surface plasmon waves with wavelength on the sub-micron scale, in contrast with standard plasmonic metals where nonlocal effects are important only when the plasmon wavelength is at the nanoscale.

The qualitative difference between the hydrodynamic model and the local Drude model persists over a wide range of loss rates $\gamma$ in Eq. \eqref{eq:v}. In Fig. \ref{fig:nonlocalmodel}(b), we plot the dispersion relation for $\gamma=0.025\omega_p$ in blue for the local model and red for the nonlocal model. The solid lines represent $\Real(K)$, while the dotted lines represent $\Imag(K)$. Within the local model, while propagation is not strictly unidirectional in the presence of loss, the counter-propagating mode is significantly overdamped ($\Real(K)<\Imag(K)$) in the unidirectional frequency range, marked by the grey dotted lines. On the other hand, the counter-propagating mode continues to remain underdamped ($\Real(K)>\Imag(K)$) in the nonlocal model. Only for substantially high values of loss ($\gamma > 0.08\omega_p$) does the dispersion relation in the nonlocal model return approximately to its local form, in which case damping is high enough that the propagation of the surface plasmon-polariton is no longer apparent. The analysis here indicates that the effect of nonlocality on nonreciprocal surface plasmon-polaritons should be pronounced for a wide range of values of loss.

\begin{figure}
		\centering
		\includegraphics[scale=0.90]{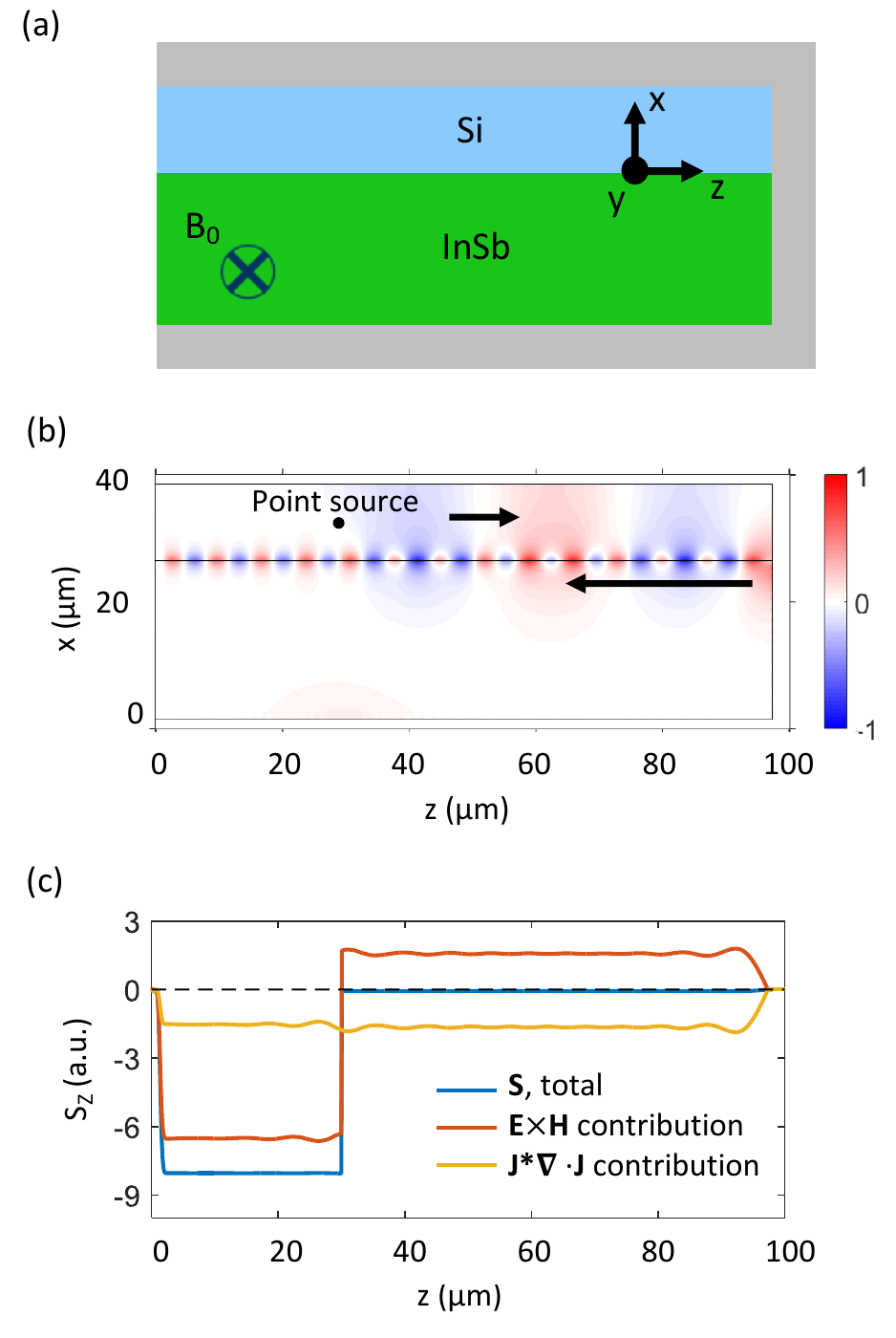}
		\caption{Numerical results using FDFD at $\omega=0.7\omega_p$. (a) The structure considered in Ref. \citep{tsakmakidisSMP}. (b) Field profile of $H_z$ generated by the indicated point source, clearly depicting a backward propagating mode with a significantly smaller wavevector than the forward propagating mode. (c) A finite Poynting flux in the backward direction relative to the point source, and a zero Poynting flux forward owing to the PEC to the right. The blue curve is the total Poynting vector from Eq. \eqref{eq:poynting}, while the red and yellow curve are its electromagnetic and kinetic terms, respectively, from Eq. \eqref{eq:poynting}.}
\label{fig:numerical}
\end{figure} 

In order to numerically demonstrate the effect of nonlocality on nonreciprocal photon transport, we re-examine the structure shown in Fig. \ref{fig:numerical}(a), which was first considered in Ref. \citep{tsakmakidisSMP}. The structure is two-dimensional and consists of the Si-InSb interface as discussed above, subject to a static out-of-plane magnetic field. Such an interface thus behaves as a nonreciprocal plasmonic waveguide. The waveguide is surrounded by a metal region, which serves both to truncate the waveguide at one end, as well as to eliminate any radiation losses. In Ref. \citep{tsakmakidisSMP}, the surrounding region was assumed to be silver. Here for simplicity we assume a surrounding region made of a perfect electric conductor (PEC), which makes very little difference to the simulations since the operating frequency, in the far-infrared region, is far below the plasma frequency of silver. Ref. \citep{tsakmakidisSMP} treats the InSb layer using the local dielectric function of Eq. \eqref{eq:tensor}. The choice of the magnetic field along the $-\hat{y}$ direction results in a unidirectional frequency range where there is a surface plasmon-polariton propagating towards the truncation, but not in the opposite direction. Consequently, at a frequency inside the unidirectional range, Ref. \citep{tsakmakidisSMP} shows that the electromagnetic field will propagate towards and be trapped at the truncation, with no leakage either in the backward direction, or through radiation losses. Such a trapping effect appears to lead to the violation of the time-bandwidth product constraint. 

On the other hand, as we have discussed above, the nonlocal behavior is in fact intrinsic and rather significant in the dielectric response of InSb. Therefore, we extend the finite-difference frequency-domain method \citep{wonseok} to include the nonlocal response as described by Eq. \eqref{eq:j} for the InSb region, and re-simulate the structure in Fig. \ref{fig:numerical}(a). To highlight the fact that there is a backward propagating mode even in the lossless system, we assume $\gamma = 0$. We excite the waveguide mode by placing a point dipole source in the vicinity of the interface. We choose $\omega= 0.7\omega_p$, a frequency that is inside the unidirectional range of the local model. In Fig. \ref{fig:numerical}(b), we plot the field distribution of $H_z$, the $z$-component of the magnetic field of the excited surface plasmon-polariton mode. We observe a significant excitation of backward propagating surface plasmon-polariton, as well as significant reflection at the truncation, in consistency with our dispersion relation analysis as shown in Fig. \ref{fig:nonlocalmodel}. 

To further highlight the contrast between the local and nonlocal models, we note that, for $\gamma=0$, the local Drude model would predict that within the unidirectional frequency range, there is a net energy flux towards the truncation, as Ref. \citep{tsakmakidisSMP} shows. On the other hand, we compute the Poynting vector flux along the $z$-direction in the nonlocal model. The time-averaged Poynting vector $\textbf{S}$ in the hydrodynamic model can be derived by combining the linearized forms of Eqs. \eqref{eq:v} and \eqref{eq:n} with the Poynting theorem to obtain
	\begin{equation}
	\textbf{S} = \frac{1}{2}\Real\Big[\E\times\Hv^* + i\frac{\beta^2}{\omega\omega_p^2\epsilon_0}\J^*(\na\cdot\J)\Big]
	\label{eq:poynting}
	\end{equation}
We show the Poynting flux along the $z$-direction in Fig. \ref{fig:numerical}(c). The total Poynting flux (blue curve) within the hydrodynamic model contains contributions from both the electromagnetic field ($\E\times\Hv^*$, red curve) and the kinetic energy of the free carriers ($\J^*\left(\na\cdot\J\right)$, yellow curve), unlike in the local model. Since there is a PEC termination, the total Poynting vector to the right of the point source must be zero. We see that this is indeed the case, with the forward propagating electromagnetic component being cancelled exactly by the counter-propagating kinetic component, an effect arising from the nonlocal term $\J^*(\na\cdot\J)$. Similarly, a negative value of Poynting flux is observed to the left of the point source, also confirming the excitation of the high-$K$ mode propagating backwards. 

\textit{General thermodynamic argument.---}In the lossless local Drude model, the surface plasmon has a dispersion relation $\omega(K)$ that asymptotically approaches a constant in the $K \to \infty$ limit. This behavior is the key to the existence of the unidirectional frequency range when a magnetic field is applied. Here we provide a simple thermodynamic argument to show that such an asymptotic behavior is nonphysical. As a result, the kind of unidirectional frequency range as predicted using the local Drude model cannot exist in any physical system. 

The argument is as follows: consider a static translationally invariant system which supports a bounded dispersion relation, i.e., $\omega(K)\in[0,\omega_0]$. Then, the number of photonic states in the frequency range $[0,\omega_0]$ is infinite, since the wavevector $K\in[0,\infty)$. Consequently, the electromagnetic energy $E$  contained in this system per unit area at temperature $T$ is 
\begin{align}
E &= \int_0^{\omega_0} \rho(\omega) \Theta(\omega,T) d\omega \\
&\ge \Theta(\omega_0,T)\int_0^{\omega_0} \rho(\omega) d\omega \\ 
&= +\infty
\end{align} 
where $\int_0^{\omega_0}\rho(\omega)=+\infty$ is the number of states per unit area in the indicated frequency range. $\Theta(\omega,T) = \hbar\omega/\left(e^{\hbar\omega/k_BT}-1\right)$ is the Bose-Einstein distribution, monotonically decreasing in $\omega$. 

Any physical system should not have infinite thermal electromagnetic energy density at a finite temperature. Thus, the dispersion relation of the lossless Drude model, and the associated undirectional frequency range, is not physical. Further, our prediction that a unidirectional frequency range does not arise when a more realistic nonlocal model is used, should therefore hold true independent of the details of the nonlocal model. 

\textit{Conclusion.---} To summarize this Letter, we show that the unidirectional frequency range, which is predicted for a free-electron metal under a static magnetic field using a local Drude model, is nonphysical. We present a general argument from thermodynamic considerations, and illustrate the argument with an explicit calculation using a more realistic nonlocal hydrodynamic model for the metal. Our results here suggest that the anomalous time-bandwidth product predicted by Ref. \citep{tsakmakidisSMP}, which arises as a direct consequence of the existence of the unidirectional frequency range, is not physical either. More generally, our work highlights the importance of using a more realistic permittivity model, such as those derived from quantum plasmonics considerations \citep{mortensenGNOR, MaierReview, jacobPlasmonicReview, estebanQuantumPlasmonics}, to understand nonreciprocal plasmonic effects. 
	
This work is supported by the AFOSR MURI projects (FA9550-18-1-0379, FA9550-17-1-0002), and by a Vannevar Bush Faculty Fellowship from the U. S. Department of Defense (1209458-601-TDAJC). S. F. acknowledges discussions with Prof. D. A. B. Miller.

\end{document}